\documentclass[journal=jacsat,manuscript=article]{achemso}

\usepackage[version=3]{mhchem} 
\usepackage{amsmath}
\usepackage{braket}
\usepackage{algorithm}
\usepackage{algpseudocode}
\usepackage{threeparttable}
\usepackage{hyperref}
\usepackage{longtable}
\usepackage{amsmath,amsfonts,amssymb}
\usepackage{comment}
\usepackage[bb=boondox]{mathalfa}
\usepackage{xr}

\author{Lukas Kim}
\affiliation{Department of Chemistry, University of California, Berkeley}
\author{Teresa Head-Gordon}
\affiliation{Department of Chemistry, University of California, Berkeley}
\email{thg@berkeley.edu}

\title[]{Near Equivalence of Polarizability and Bond Order Flux Metrics for Describing Covalent Bond Rearrangements}


\begin{document}

\begin{tocentry}
    \centering
    \includegraphics[width=1\linewidth]{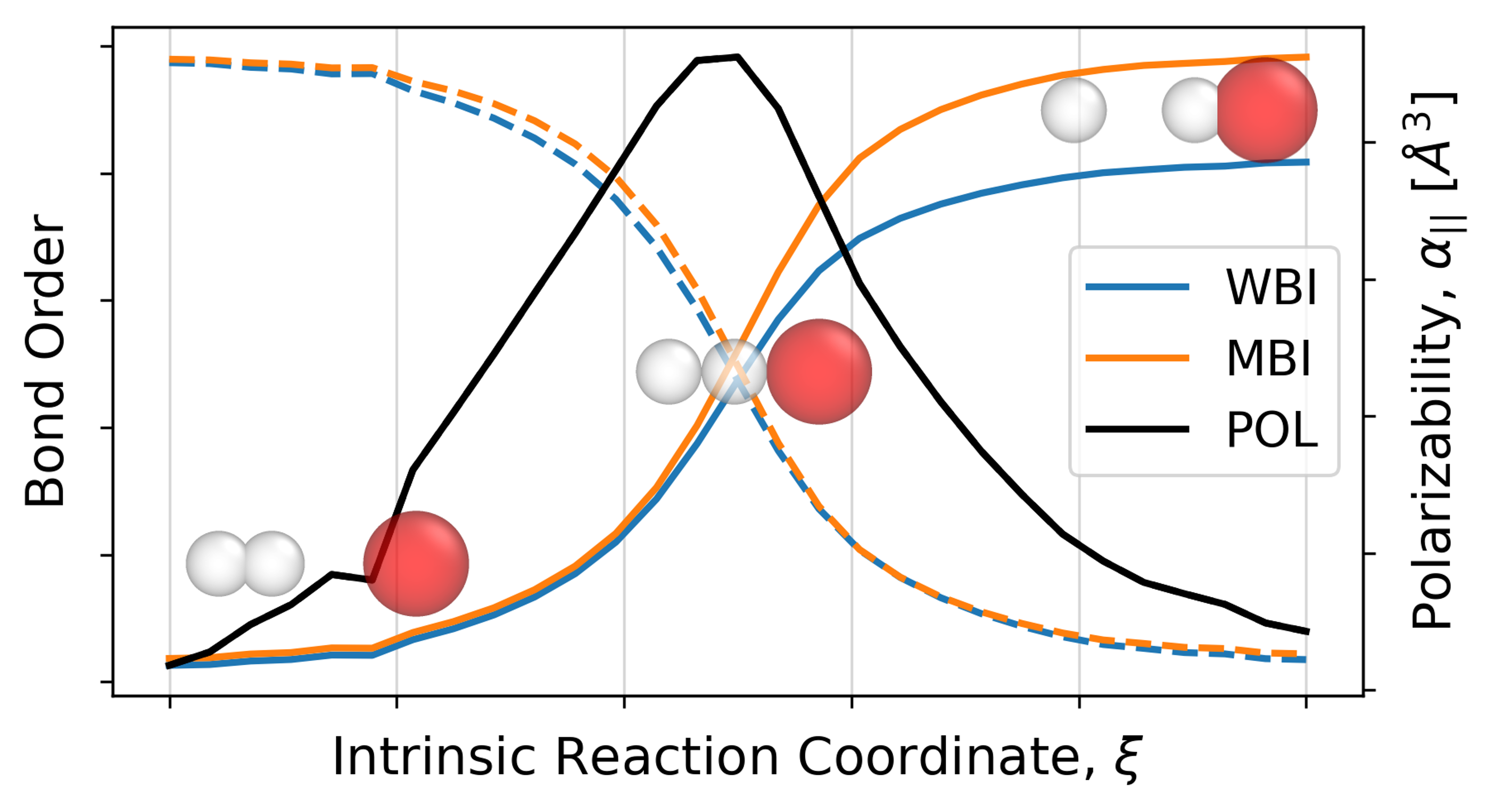}
\end{tocentry}

\begin{abstract}
\noindent
Identification of the breaking point for the chemical bond is essential for our understanding of chemical reactivity. The current consensus is that a point of maximal electron delocalization along the bonding axis separates the different bonding regimes of reactants and products. This maximum transition point has been investigated previously through the total position spread and the bond-parallel components of the static polarizability tensor for describing covalent bond breaking. In this paper, we report that the first-order change of the Wiberg and Mayer bond index with respect to the reaction coordinate, the bond flux, is similarly maximized and is nearly equivalent with the bond breaking points determined by the bond-parallel polarizability. We investigate the similarites and differences between the two bonding metrics for breaking the nitrogen triple bond, twisting around the ethene double bond, and a set of prototypical reactions in the hydrogen combustion reaction network. The Wiberg-Mayer bond flux provides a simpler approach to calculating the point of bond dissociation and formation and can yield greater chemical insight through bond specific information for certain reactions where multiple bond changes are operative.
\end{abstract}

\section{Introduction}
Stable molecules are defined by their unique arrangement of chemical bonds.\cite{Ruedenberg1962} Quantum mechanical (QM) calculations can provide information about the energetics and electron density of molecules using methods that can often reach high accuracy, although it is difficult to conceptualize the chemical bonding within the abstraction of wave function or density functional theory formulations.\cite{Outeiral2018,Krylov2020} Hence methods of translating the results of QM computations into the vernacular of a chemical theory of bonding are broadly referred to as wave function analysis methods. The overarching goal of these methods is to provide better connections between QM definitions and conceptual chemical properties, such as bond order, atomic charge, and electronegativity in order to understand stable molecules. Different classes of interpretative chemical tools include the quantum theory of atoms in molecules (QTAIM) framework of Bader\cite{Bader1974}, natural bond order analysis of Weinhold\cite{Weinhold2012}, and the Wiberg and Mayer bond indices that have been used qualitatively for many decades to characterize stable molecular topologies. \cite{wiberg_application_1968, mayer_bond_2007}

During a chemical reaction, however, the bonds of a reactant become partially broken and/or new bonds form at a transition region that ultimately progresses to a new arrangement of stable chemical bonds in a product molecule. The definition of the breaking point of the chemical bond is central to the mechanistic interpretation of chemical reactions, and is still an open question in the theory of chemical bonding and wave function interpretative tools. It is known that electron delocalization and localization are critical indicators of bond (de)formation in chemical reactions.\cite{GalloBueno2016, Brea2013} To illustrate, consider a concerted substitution reaction mechanism in which a bond is simultaneously broken between fragments $\text{AB}$ and formed between fragments $\text{BC}$,
\begin{equation}
    \ce{A}\text{-}\ce{B} + \ce{C}
    \Leftrightarrow
    [\ce{A} \text{---} \ce{B} \text{---} \ce{C}]^{\ddagger}
    \Leftrightarrow
    \ce{A} + \ce{B}\text{-}\ce{C}
\end{equation}
The reaction proceeds through a transition state region where the fragments $\text{A}$ and $\text{C}$ are both partially bonded to the transferred fragment $\text{B}$. At the transition state, the electron density is maximally delocalized due to the elongated and partially formed bonds spanning across $\text{ABC}$, in contrast to the localized reactant and product states. 

Computationally, one principal measure of electron (de)localization is the total position spread (TPS) tensor, an analytical measure of the spread of the electron density. In short, it is the variance in the sum total of the electron positions, a property that has been shown to have a maximum along bond breaking reaction coordinates.\cite{Brea2013} Recently, Hait and M. Head-Gordon showed that the static polarizability has a similar maximum along bond dissociation coordinates marking the breaking (or forming) of a chemical bond.\cite{hait_when_2023} For context, the static or dipole polarizability tensor relates the induced dipole $\vec{\mathbf{p}}$ of a molecule as proportional to an applied electric field $\vec{\mathbf{E}}$ 
\begin{equation}
    \vec{ \mathbf{p} } = \overset{\leftrightarrow}{\alpha} \cdot \vec{\mathbf{E}}, \qquad \text{where} \quad {\alpha}_{ij} = \left(\frac{\partial p_i}{\partial E_j}\right) 
\end{equation}
where $i$ and $j$ index the Cartesian axes. In our example reaction above, the parallel polarizability, i.e. along the bond axis, is expected to be maximized at the transition state since a perturbing electric field will bias bond formation in one direction and naturally have a large effect on the displacement of the electron density, moving electrons located in the partially formed $\text{AB}$ bond to $\text{BC}$ bond. This maximum has been shown to appear in both homolytic and heterolytic bond dissociations, although polarizability is less descriptive when analyzing $\pi$-bond rotations such as for ethene. \cite{hait_when_2023} 

The polarization metric is directly proportional to the TPS tensor along the reaction coordinate, but is further augmented by a denominator quantity that describes the gap between bonding
and anti-bonding orbitals, a quantity that is minimized for more polarizable bonds.\cite{hait_when_2023} Furthermore, macroscopic polarization and electron localization have been found to be intimately related in the study of insulating and conducting materials.\cite{Resta2002} In this paper, we report that a maximum in the first order derivative of the Mayer/Wiberg orbital-based bond indices show excellent correspondence with the bond breaking point of the static polarizability maximum. The bond breaking point is characterized by maximal sensitivity of the bond order to displacement and marks an inflection point, separating the convex and concave regions of the bond order along the reaction coordinate. We show that the Wiberg and Mayer bond order derivatives are quite robust across multiple reaction channels for hydrogen combustion, diatomic nitrogen dissociation, and twisting around the ethene bond.

\section{Theory and Methods}
\textbf{Wiberg and Mayer Bond Indices.} Within QTAIM, it has been shown that two-body decomposition of the TPS tensor defines the two-body delocalization index (DI), which relate to the formal bond orders used for stable molecules.\cite{Outeiral2018} Furthermore, the DI has been shown to be a real space analogue of the orbital-based Wiberg/Mayer bond indices, which we derive in Appendix A.\cite{Bader1974, Outeiral2018} 

The Mayer and Wiberg bond indices are measures of bond order and are computed from the first order reduced density (1-RDM) and the orbital overlap matrices. In an atomic orbital basis, the bond indices are obtained by summing the block-off-diagonal components, corresponding to a sum of the overlap density of each orbital pair between atom centers.\cite{mayer_bond_2007, wiberg_application_1968}
\begin{equation}
    \text{MBI}_{AB} = 2\sum_{\mu \in A} \sum_{\nu \in B} 
    \left[(P^\alpha S)_{\mu\nu} P^\alpha S)_{\nu\mu} + (P^\beta S)_{\mu\nu}(P^\beta S)_{\nu\mu}\right]
\end{equation}
\begin{equation}
    \text{WBI}_{AB} = 2\sum_{\mu \in A} \sum_{\nu \in B} 
    \left|P^\alpha_{\mu\nu} + P^\beta_{\mu\nu}\right|^2
\end{equation}
In this way, these simple metrics quantify the number of electrons ‘shared’ by two atom centers, analogous to the bonding concepts of classical valence bond theory. 

Here we will consider the change in the bond order with respect to the intrinsic reaction coordinate (IRC) for the various hydrogen and oxygen transfer reactions in the hydrogen combustion reaction network. The bond order flux is defined as the derivative of the bond index with respect to a reaction coordinate.
\begin{equation}
    J_{b} = \frac{\partial \text{BI}_{AB}}{\partial \xi_{\text{IRC}}}
    \label{eq:flux}
\end{equation}
Our primary objective is to show that Eq. \ref{eq:flux} is in good agreement with the polarization metric, provides a simpler approach to calculating the point of bond dissociation and formation, and yields greater chemical insight into the hydrogen combustion reactions.

\textbf{Hydrogen combustion data}. The polarizability, bond orders, and bond flux metrics were investigated for 13 reaction channels of hydrogen combustion as a model reactive system.\cite{guan_benchmark_2022, li_updated_2004}
The set of reactions contains hydrogen and oxygen transfer reactions, substitution, and diatomic or bimolecular bond dissociation profiles. The original dataset contains geometries, energies, and forces from intrinsic reaction coordinate (IRC) scans, \textit{ab initio} molecular dynamics, and normal mode displacements. 

\vspace{5mm}

\begin{table}
    \begin{tabular}{|c|l|} \hline 
        Index&  Reaction\\ \hline 
        \multicolumn{2}{|l|}{\textbf{Substitution}}\\ \hline 
        16& H\textsubscript{2}O\textsubscript{2} + H $ \rightarrow $ H\textsubscript{2}O + OH\\ \hline 
        \multicolumn{2}{|l|}{\textbf{Oxygen Transfer}}\\ \hline 
        1 &  H + O\textsubscript{2} $\rightarrow$ OH + O\\ \hline 
        11&  HO\textsubscript{2} + H $\rightarrow$ 2OH\\ \hline 
        12&  HO\textsubscript{2} + O $\rightarrow$ OH + O\\ \hline 
        \multicolumn{2}{|l|}{\textbf{Hydrogen Transfer}}\\ \hline 
        2& O + H\textsubscript{2} $\rightarrow$ OH + H\\ \hline 
        3& H\textsubscript{2} + OH $\rightarrow$ H\textsubscript{2} O + H\\ \hline 
        4& H\textsubscript{2} O + O $\rightarrow$ 2OH\\ \hline 
        10&HO\textsubscript{2} + H $\rightarrow$ H\textsubscript{2} + O\textsubscript{2}\\ \hline 
        13&HO\textsubscript{2} + OH $\rightarrow$ H\textsubscript{2} O + O\textsubscript{2}\\ \hline 
        14&2HO\textsubscript{2} $\rightarrow$ H\textsubscript{2}O\textsubscript{2} + O\textsubscript{2}\\ \hline 
        17&H\textsubscript{2}O\textsubscript{2} + H $\rightarrow$ HO\textsubscript{2} + H\textsubscript{2}\\ \hline 
        18&H\textsubscript{2}O\textsubscript{2} + O $\rightarrow$ HO\textsubscript{2} + OH\\ \hline 
        19&H\textsubscript{2}O\textsubscript{2} + OH $\rightarrow$ H\textsubscript{2}O + HO\textsubscript{2}\\ \hline
    \end{tabular}
    \caption{\textit{Selected reactions in the kinetic model of hydrogen combustion investigated in this study.} The IRC data developed in references [\cite{guan_benchmark_2022} and \cite{Bertels2020}] were analyzed with the parallel bond-projected polarizability and bond order and bond flux metrics.}
    \label{tab:rxn-table}
\end{table}

\textbf{Computational Details.} Bond indices, bond flux, and static polarizabilities were computed for the 13 reaction channels of hydrogen combustion using the NBO 5.0 Program \cite{glendening_nbo_nodate} integrated into Q-Chem version 5.4 \cite{epifanovsky_software_2021}. Optimized geometries along the intrinsic reaction coordinate (IRC) were obtained from the benchmark dataset for hydrogen combustion.\cite{guan_benchmark_2022} Electronic structure calculations were carried out at the DFT level, with consistent functional and basis set used to generate the geometries in the dataset, namely the $\omega$B97x-V density functional \cite{Mardirossian2014} and Dunning’s triple-zeta correlation-consistent basis set.\cite{dunning_gaussian_1989} The bond order, bond order flux, and polarizability profiles were computed with spin-unrestricted CASSCF \cite{Roos1980, epifanovsky_software_2021} using a (6,6) and (2,2) active space for nitrogen and ethene, respectively.

In order to imbue the static polarizability with a degree of pair-specificity, we define a bond-projected polarizability to study the concerted rupture and formation of multiple bonds.
\begin{equation}
    \alpha_{AB} = (\overset{\leftrightarrow}{\alpha} \cdot \hat{r}_{AB}) \cdot \hat{r}_{AB}
\end{equation}
This quantity is the projection of the induced molecular dipole in the direction along the bond for a unit applied electric field in its direction. For a linear transition state, the projections in the bond-breaking and bond-forming directions will be identical and yield identical results to the original, axis-aligned polarizability metric. By comparing the component of the molecular polarizability tensor along each bond dissociation or bond association coordinate, we can assess the degree of non-linearity of the transition state. 

\section{Results and Discussion}

The hydrogen and oxygen transfer reactions in Table \ref{tab:rxn-table} all proceed through a transition state geometry where the transferred atom is partially bonded to both molecular fragments. Supplementary Figure 1(a) summarizes the bond-projected polarizability and Mayer and Wiberg bond indices along the intrinsic reaction coordinate for all 13 hydrogen combustion reactions summarized in Table \ref{tab:rxn-table}. In each case, the position at which the polarizability peaks along the reaction coordinate describes the broken bond transition point, while the Wiberg and Mayer bond order crossover points are where the slope along the reaction coordinate is maximized. Hence by taking the derivative of the bond index with respect to the reaction coordinate, the correspondence of the change in the bond order with the bond-projected polarizability becomes even more evident as seen in Supplementary Figure 1(b) for all hydrogen combustion reactions.  In what follows, we discuss 3 of the 13 reactions, each demonstrating the ability of the bond order and flux to discern unique bond rearrangement cases that are in excellent agreement with the polarizability metric, while also offering chemical insight.

\begin{figure}
    \centering
    \includegraphics[width=1\linewidth]{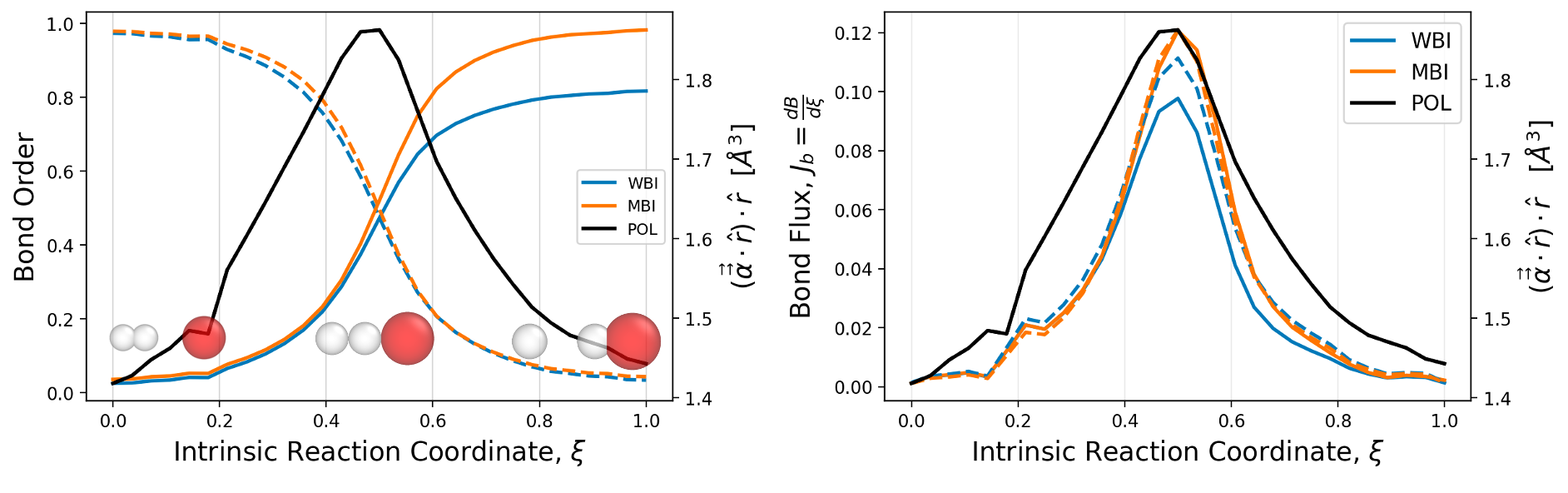}
    \caption{\textit{A prototypical example of a $\sigma$-to-$\sigma$ bond transfer reaction.} Shown for reaction 2, a hydrogen transfer reaction. (a) Bond order and (b) bond order flux for the Wiberg (blue) and Mayer (orange) bond indices plotted alongside the bond projected polarizability (black). The plots are overlaid and are plotted on separate scales. Solid and broken lines indicate forming and breaking bonds, respectively.}
    \label{fig:RXN02}
\end{figure}

In Figure \ref{fig:RXN02}, the bond order and bond order flux profiles are plotted for hydrogen transfer reaction 2, in which the hydrogen-hydrogen $\sigma$-bond is broken and a hydrogen-oxygen $\sigma$-bond is formed. This $\sigma$-to-$\sigma$ transfer reaction is the simplest non-degenerate bond rearrangement in hydrogen combustion. The bond-projected polarizability peak correlates with the bond order crossover point at which the bond order takes a value of one half (Figure \ref{fig:RXN02}a). As expected, the transition state is characterized by equal partitioning of the transferred atom's valence, resulting in two partial bonds that span the triplet of atoms. As the polarizability indicates the total electron delocalization (i.e. across all three atoms), the electron density is maximally delocalized at the transition state. On the other hand, the bond indices indicate the extent of pairwise electron delocalization across specific bonds, and the maximum of the bond flux at the transition state indicates the instability point of the electron delocalization that is in good agreement with the polarizability (Figure \ref{fig:RXN02}b).

\begin{figure}
    \centering
    \includegraphics[width=1\linewidth]{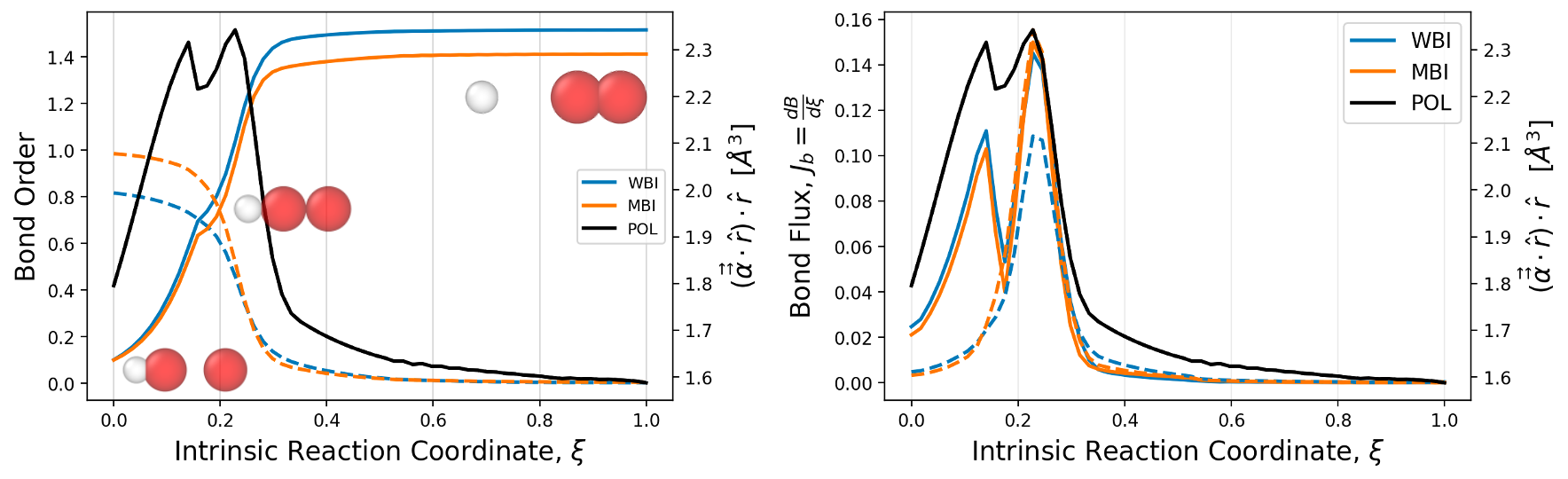}
    \caption{\textit{A prototypical example of a reaction revealing two bond formation steps.} Shown for reaction 1, a oxygen transfer reaction. (a) Bond order and (b) bond order flux for the Wiberg (blue) and Mayer (orange) bond indices plotted alongside the bond projected polarizability (black). The plots are overlaid and are plotted on separate scales. Solid and broken lines indicate forming and breaking bonds, respectively.}
    \label{fig:RXN01}
\end{figure}

However, the reaction 1 oxygen transfer case in Figure \ref{fig:RXN01} shows that two peaks are observed in the parallel polarizability profile, with the oxygen-oxygen sigma bond forming first, indicated by the initial rise in the oxygen-oxygen bond order, while the oxygen-hydrogen bond is relatively undisturbed. The second peak in the polarizability is due to the $\sigma$-to-$\pi$ bond rearrangement, forming a triplet oxygen molecule and a lone hydrogen atom. The Mayer/Wiberg bond indices correctly capture the bond order of 1.5 for the triplet oxygen molecule due to its pair of two-center three-electron (2c-3e) bonds. This step-wise progression would be missing from an energetic perspective, since no stable intermediate is formed at the first peak. The transition state lies on the second peak corresponding to the $\sigma$-to-$\pi$ bond rearrangement.

\begin{figure}
    \centering
    \includegraphics[width=1\linewidth]{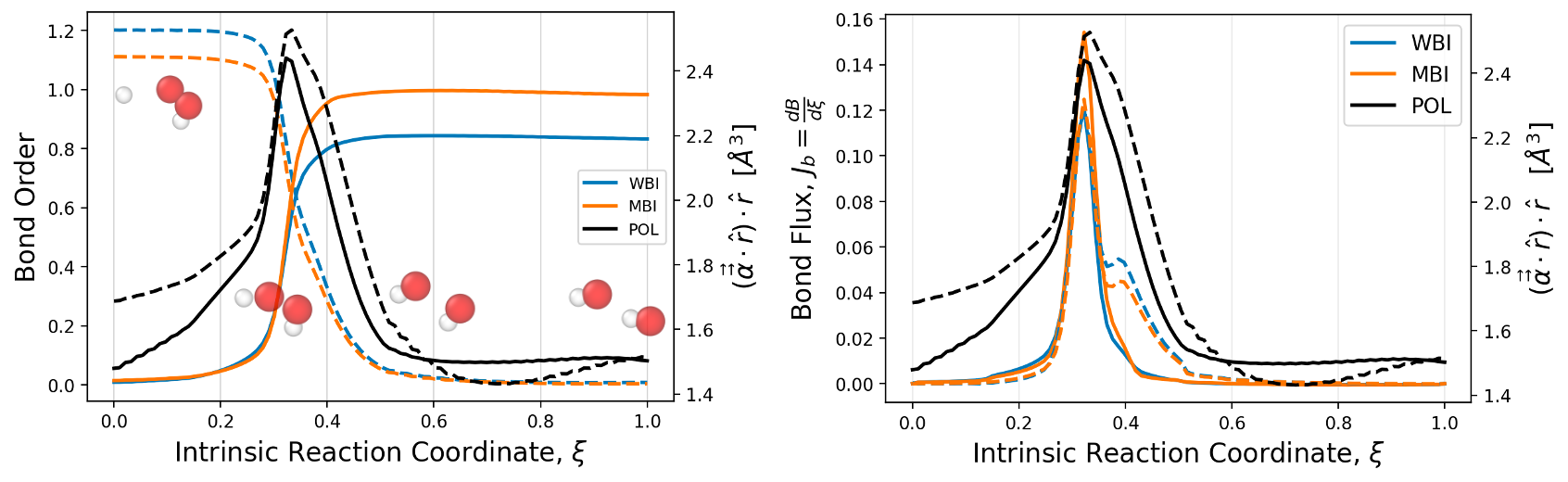}
    \caption{\textit{A prototypical example of a $\sigma$-to-$\sigma$ and $\sigma$-to-$\pi$ bond transfer reaction.} Shown for reaction 11, a oxygen transfer reaction. (a) Bond order and (b) bond order flux for the Wiberg (blue) and Mayer (orange) bond indices plotted alongside the bond projected polarizability (black). The plots are overlaid and are plotted on separate scales. Solid and broken lines indicate forming and breaking bonds, respectively.}
    \label{fig:RXN11}
\end{figure}

The reaction 11 oxygen transfer in Figure \ref{fig:RXN11} corresponds to a case where the bond flux finds a step-wise formation of oxygen-oxygen $\sigma$- and $\pi$-bonds, whereas as the parallel polarizability profile is more ill-defined. The bond order profile seems to proceed via a $\sigma$-to-$\sigma$ bond transfer from the oxygen-oxygen to oxygen-hydrogen bond as seen in Figure \ref{fig:RXN11}(a), but unlike the previous examples which were linear rearrangements, the perpendicular contributions from other bonds now play a role. This gives rise to a small peak in the bond flux profile along the breaking oxygen-oxygen bond as seen in Figure \ref{fig:RXN11}(b). This increase has little effect on the OH bond order but manifests as an increase in the oxygen-oxygen bond order. This increase is attributed to the occupied-virtual orbital interaction, or "charge transfer" interaction, between the oxygen lone-pair and the newly formed OH anti-bonding orbital. In the reverse reaction, the increase in bond order is comparable to an activation of the OH bond and is a key feature of the minimum energy reaction pathway.

\begin{figure}
    \centering
    \includegraphics[width=1\linewidth]{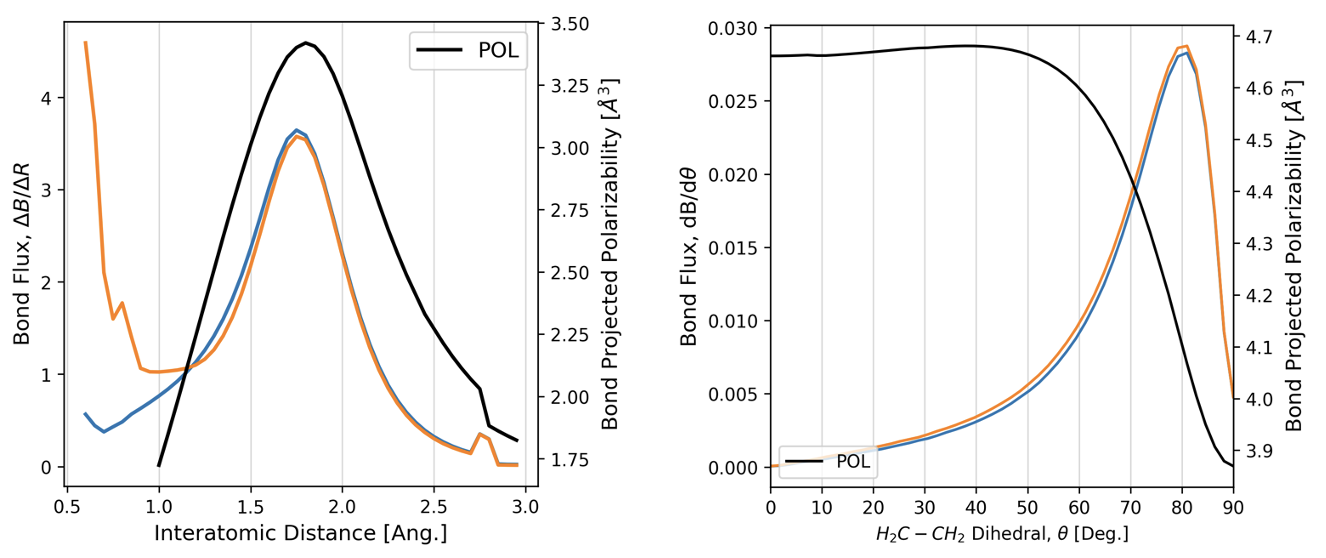}
    \caption{\textit{Nitrogen molecule dissociation and rotation around the ethene double bond.} Bond flux profiles for the Wiberg (WBI) and Mayer (MBI) bond indices, plotted against the bond-projected polarizability (a) computed with spin-unrestricted CASSCF(6,6)/cc-pVTZ for nitrogen dissociation and (b) computed with spin-unrestricted CASSCF(2,2)/cc-pVTZ for rotation around the ethene C=C bond.}
    \label{fig:nitrogen-dissociation}
\end{figure}

As reported by Hait and M. Head-Gordon, two peaks for $\sigma$- and $\pi$-bond rupture were not observed for the diatomic dissociation of the nitrogen molecule, attributed to inadequate separation of the length scales for the breaking points of the $\sigma$- and $\pi$-bonds.\cite{hait_when_2023} In other words, the $\sigma$- and $\pi$-bonds break nearly simultaneously as the two fragments are pulled apart and not in a step-wise fashion with the $\pi$-bonds breaking first and the $\sigma$-bond last. We find this is also the case for the WBI/MBI bond order and bond flux profiles (Figure \ref{fig:nitrogen-dissociation}a and Supplementary Figure 2a). The $\sigma$- and $\pi$-bonds break simultaneously with an inflection point at a bond order of half its equilibrium value.

Another purported instance of inadequate length scale separation is the breaking of the ethene $\pi$-bond by rotation (Figure \ref{fig:nitrogen-dissociation}b and and Supplementary Figure 2b). In this case, the polarizability possesses only a small peak due to the small spatial separation of the radical fragments. However, the bond flux profile predicts a bond breaking point near the 80 degree dihedral rotation. The lack of spatial separation is not an issue for the WBI/MBI and the $\pi$-bond breaks as expected due to the lack of p-orbital overlap density, i.e. the restrictions of atomic orbital symmetry. While the Mayer and Wiberg bond indices have a close relationship to the polarizability via the total position spread tensor, the bond indices capture the electronic structure in the space of atomic orbitals rather than real space. This is a particular advantage in such cases where spatial isolation of the radical fragments is small.

\section{Conclusion}
In summary, two chemical concepts, the dipole polarizability observable from quantum mechanics and the bond flux metric derived as a derivative quantity of bond order indices from wavefunction analysis, are shown to be directly related quantities for resolving a chemically intuitive picture of continuous bond rearrangements. We have shown in explicit examples that the sensitivity of bond order to displacements, i.e. bond flux, is correspondingly maximized along with the polarizability. By definition, the covalent bond order is a pairwise measure of electron delocalization, i.e. it quantifies the number of electrons shared by a pair of atom centers. Similarly, the close mathematical relationship between the polarizability, which is measurable in principle, and the total position spread (TPS) tensor facilitates its connection to molecular electron delocalization. The TPS is similarly related to the Mayer/Wiberg indices,\cite{Outeiral2018} indicating that polarizability and the bond order flux offer near equivalence in many bonding scenarios. 

Since orbital bond indices are derived from the one-particle reduced density matrix, the bond indices are computed with negligible additional cost. Orbital-based bond indices also have additional advantages such as bond specific information and that they are not restricted to linear dissociative reaction coordinates, as illustrated for rotation around the ethene double bond. At the same time, we note that polarizability has a distinct advantage by being able to describe bond breaking involving avoided crossings for some homolytic bonds.\cite{hait_when_2023}

\section{Appendix: Exchange Density and Mayer Bond Index}
The definition of the exchange density follows from the decomposition of the expectation value of the pair density operator $\hat{\rho}_2(\vec{r}_1, \vec{r}_2)$,
\begin{equation}
    \hat{\rho}_2(\vec{r}_1, \vec{r}_2)
    = \sum_{i<j} \left[
    \delta(\vec{r}_i - \vec{r}_1)\delta(\vec{r}_j - \vec{r}_2)
    + \delta(\vec{r}_j - \vec{r}_1)\delta(\vec{r}_i - \vec{r}_2)
    \right]
\end{equation}
where $i$ and $j$ run over individual electrons. Similar to how the expectation of the density operator, $\hat{\rho}(\vec{r})$, yields the probability of finding an electron at a point $\vec{r}$, the expectation of the pair density operator yields the probability of finding an electron at point $\vec{r}_1$ and simultaneously another electron at point $\vec{r}_2$. For a single determinant wave function $\Psi$ built from orthonormalized spin-orbitals $\psi_i(\vec{r}, \sigma) = \phi_i(\vec{r})\gamma_i(\sigma)$, we arrive at the following after expansion:
\begin{equation}
    \rho_2(\vec{r}_1, \vec{r}_2)
    = \braket{\Psi|\hat{\rho}_2(\vec{r}_1, \vec{r}_2)|\Psi}
    = \sum_{i,j=1}^N \left(
    \left| \phi_i(\vec{r}_1) \right|^2 \left| \phi_j(\vec{r}_2) \right|^2
    - \phi_i^*(\vec{r}_1) \phi_j(\vec{r}_1) \phi_j^*(\vec{r}_2) \phi_i(\vec{r}_2)
    \delta_{\gamma_i \gamma_j}
    \right)
\end{equation}
Since the density $\rho(\vec{r}) = \sum_i^N \left| \phi_i(\vec{r}) \right|^2$, the definition of the exchange density is given by
\begin{equation}
\begin{aligned}
\rho_2(\vec{r}_1, \vec{r}_2) &= 
    \left(\sum_{i=1}^N \left| \phi_i(\vec{r}_1) \right|^2 \right)
    \left(\sum_{j=1}^N \left| \phi_j(\vec{r}_2) \right|^2 \right)
    - \sum_{i,j=1}^N \phi_i^*(\vec{r}_1) \phi_j(\vec{r}_1) \phi_j^*(\vec{r}_2) \phi_i(\vec{r}_2)\\
    &= \rho(\vec{r}_1)\rho(\vec{r}_2) - \rho_X(\vec{r}_1, \vec{r}_2)
\end{aligned}
\end{equation}
\begin{equation}
    \rho_X(\vec{r}_1, \vec{r}_2)
    = \sum_{i,j=1}^N \phi_i^*(\vec{r}_1) \phi_j(\vec{r}_1) \phi_j^*(\vec{r}_2) \phi_i(\vec{r}_2) \delta_{\gamma_i \gamma_j}
\end{equation}
In this way, the exchange density is the 'correction' to the product of single particle probabilities due to exchange correlation. These corrections account for the so-called Fermi heap or Fermi hole, the respective increase or decrease in electron density in the vicinity of another electron due to exchange symmetry.

Using the following identities and the definition of exchange density from above, 
\begin{equation}
    n = \iint_\textit{all space} \rho_X(\vec{r}_1, \vec{r}_2) d\vec{r}_1 d\vec{r}_2
    \quad\quad\quad
    S_{\mu\nu} = \int \chi_\mu^*(\vec{r}) \chi_\nu(\vec{r}) d\vec{r}
    \quad\quad\quad
    P_{\mu\nu} = \sum_{i=1}^N c_{\mu}^i c_{\nu}^{i*}
\end{equation}
After the LCAO expansion of each MO, $\phi_i(\vec{r}) = \sum_{\mu} c_\mu^i\chi_\mu(\vec{r})$, and integration over $\vec{r}_1$ and $\vec{r}_2$,
\begin{equation}
\begin{aligned}
    n &= \sum_{i,j=1}^N \sum_{\mu, \nu, \gamma, \sigma=1}^m
    \left(\int c_\mu^{i*} c_\nu^j \chi_\mu^*(\vec{r}_1) \chi_\nu(\vec{r}_1) d\vec{r}_1\right)
    \left(\int c_\gamma^{j*} c_\sigma^{i} \chi_\gamma^*(\vec{r}_2) \chi_\sigma(\vec{r}_2) d\vec{r}_2\right) \delta_{\gamma_i \gamma_j} \\
    &= \sum_{i,j=1}^N \sum_{\mu, \nu, \gamma, \sigma=1}^m
    c_\sigma^{i} c_\mu^{i*}
    \left(\int \chi_\mu^*(\vec{r}_1) \chi_\nu(\vec{r}_1) d\vec{r}_1\right)
    c_\nu^j c_\gamma^{j*} 
    \left(\int \chi_\gamma^*(\vec{r}_2) \chi_\sigma(\vec{r}_2) d\vec{r}_2\right) 
    \delta_{\gamma_i \gamma_j}\\
    &= \sum_{i,j=1}^N \sum_{\mu, \nu, \gamma, \sigma=1}^m
    c_\sigma^{i} c_\mu^{i*}
    S_{\mu\nu}
    c_\nu^j c_\gamma^{j*} 
    S_{\gamma \sigma}
    \delta_{\gamma_i \gamma_j}\\
    &= \sum_{\nu, \sigma=1}^m \sum_{\mu, \gamma=1}^m
    P^{\alpha}_{\sigma \mu} S_{\mu\nu}
    P^{\alpha}_{\nu\gamma} S_{\gamma\sigma}
    +
    P^{\beta}_{\sigma \mu} S_{\mu\nu}
    P^{\beta}_{\nu\gamma} S_{\gamma\sigma}\\
    &= \sum_{\nu,\sigma=1}^m
    \left[
    (\textbf{P}^\alpha\textbf{S})_{\sigma\nu} (\textbf{P}^\alpha\textbf{S})_{\nu\sigma}+ 
    (\textbf{P}^\beta\textbf{S})_{\sigma\nu} (\textbf{P}^\beta\textbf{S})_{\nu\sigma}
    \right] \\
\end{aligned}
\end{equation}
Therefore, the integral of the exchange density is the total of Mayer orbital bond indices.
\begin{equation}
    \iint \rho_X(\vec{r}_1, \vec{r}_2) d\vec{r}_1 d\vec{r}_2
    = 
    \sum_{\nu,\sigma=1}^m
    \left[
    (\textbf{P}^\alpha\textbf{S})_{\sigma\nu} (\textbf{P}^\alpha\textbf{S})_{\nu\sigma}+ 
    (\textbf{P}^\beta\textbf{S})_{\sigma\nu} (\textbf{P}^\beta\textbf{S})_{\nu\sigma}
    \right]
\end{equation}
The quantity inside the summation is the orbital bond index between atomic orbitals, $\chi_\sigma$ and $\chi_\mu$. If the orbital-bond index contributions are collected by their corresponding atom centers, then the Mayer bond index contains the two-body contribution to the integral of the exchange density. We obtain the total bond order equation.
\begin{equation}
    n = \sum_{A,B} \sum_{\mu \in A} \sum_{\nu \in B}
    \left[
    (\textbf{P}^\alpha\textbf{S})_{\mu\nu} (\textbf{P}^\alpha\textbf{S})_{\nu\mu}+ 
    (\textbf{P}^\beta\textbf{S})_{\mu\nu} (\textbf{P}^\beta\textbf{S})_{\nu\mu}
    \right]
     = \frac{1}{2} \sum_{A} \text{B}_{AA} + \sum_{A \neq B} \text{B}_{AB}
\end{equation}
where $A$ and $B$ index atom centers, $\mu$ and $\nu$ index atomic orbitals centered on $A$ and $B$, respectively, and the bond order $B_{AB}$ is given by
\begin{equation}
    B_{AB} = 2 \sum_{\mu \in A} \sum_{\nu \in B}
    \left[
    (\textbf{P}^\alpha\textbf{S})_{\mu\nu} (\textbf{P}^\alpha\textbf{S})_{\nu\mu}+ 
    (\textbf{P}^\beta\textbf{S})_{\mu\nu} (\textbf{P}^\beta\textbf{S})_{\nu\mu}
    \right]
\end{equation}
The localization and delocalization indices from density-based QTAIM methods are partitioned similarly into a sum over a set of atomic basins, $\{\Omega_A\}$. \cite{Outeiral2018}
\begin{equation}
\begin{aligned}
  n 
  &= \iint_{\textit{all space}} \rho_{x}(\vec{r}_1, \vec{r}_2) d\vec{r}_1 d\vec{r}_2\\
  &= 2\sum_{A,B} \iint_{\Omega_A, \Omega_B} \rho_{x}(\vec{r}_1, \vec{r}_2) d\vec{r}_1 d\vec{r}_2\\
  &= \frac{1}{2} \sum_{A} \delta(A, A) + \sum_{B \neq A} \delta(A, B)\\
\end{aligned}
\end{equation}
where $\delta(A, B) = 2\iint_{\Omega_A, \Omega_B}\rho_{x}(\vec{r}_1, \vec{r}_2) d\vec{r}_1 d\vec{r}_2$. In this way, the QTAIM delocalization index (DI) is the two-body contribution to the integral of the exchange density, obtained through the sixth-order integration over a real space partitioning of the electron density. Formally, we can see that this can be thought of as the real space analogue of the Mayer bond index.

\begin{acknowledgement}
We acknowledge support from the U.S. National Science Foundation through Grant No. CHE-2313791.
\end{acknowledgement}

\bibliography{references}
\end{document}


\begin{figure}
    \centering
    \includegraphics[width=1\linewidth]{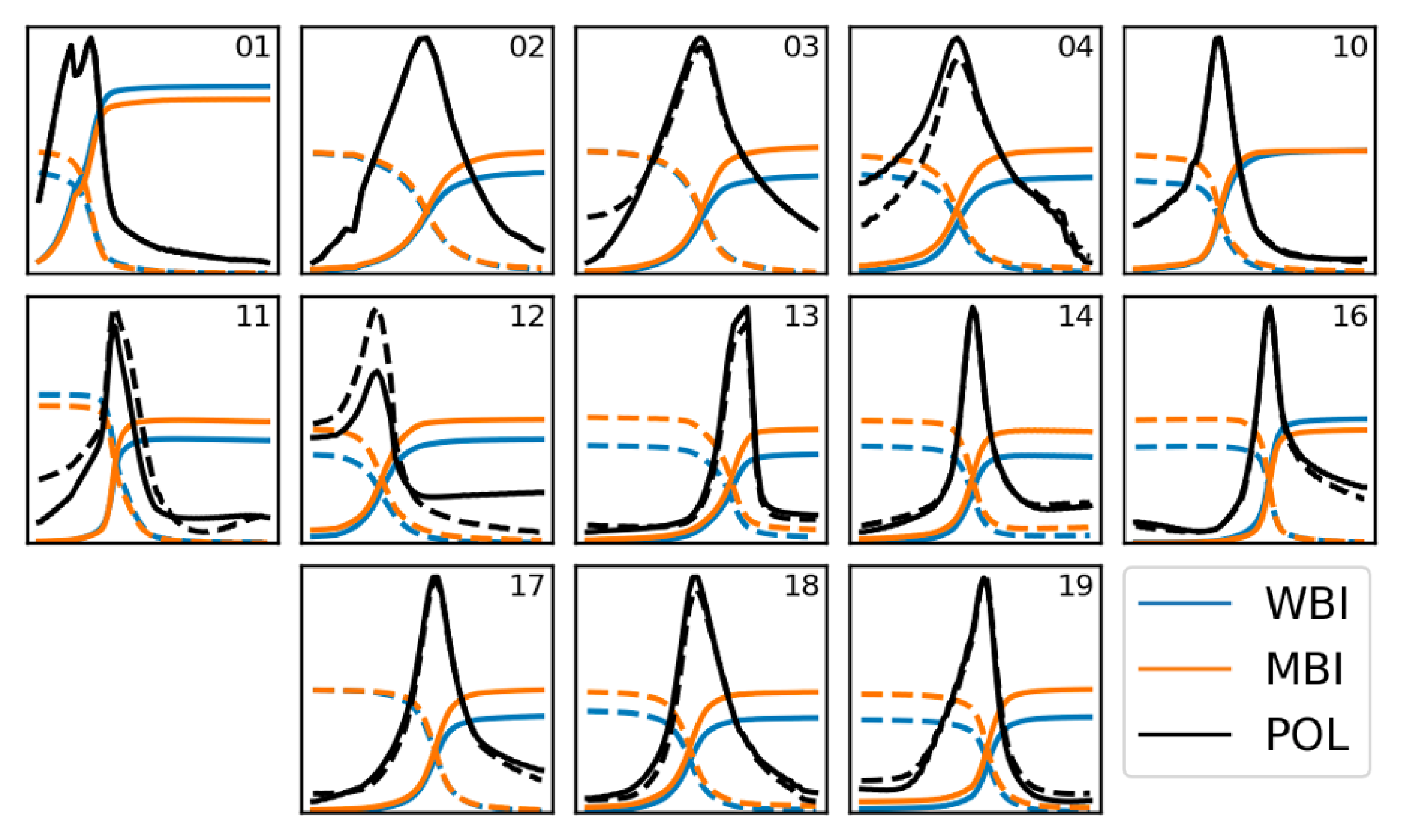}
    \includegraphics[width=1\linewidth]{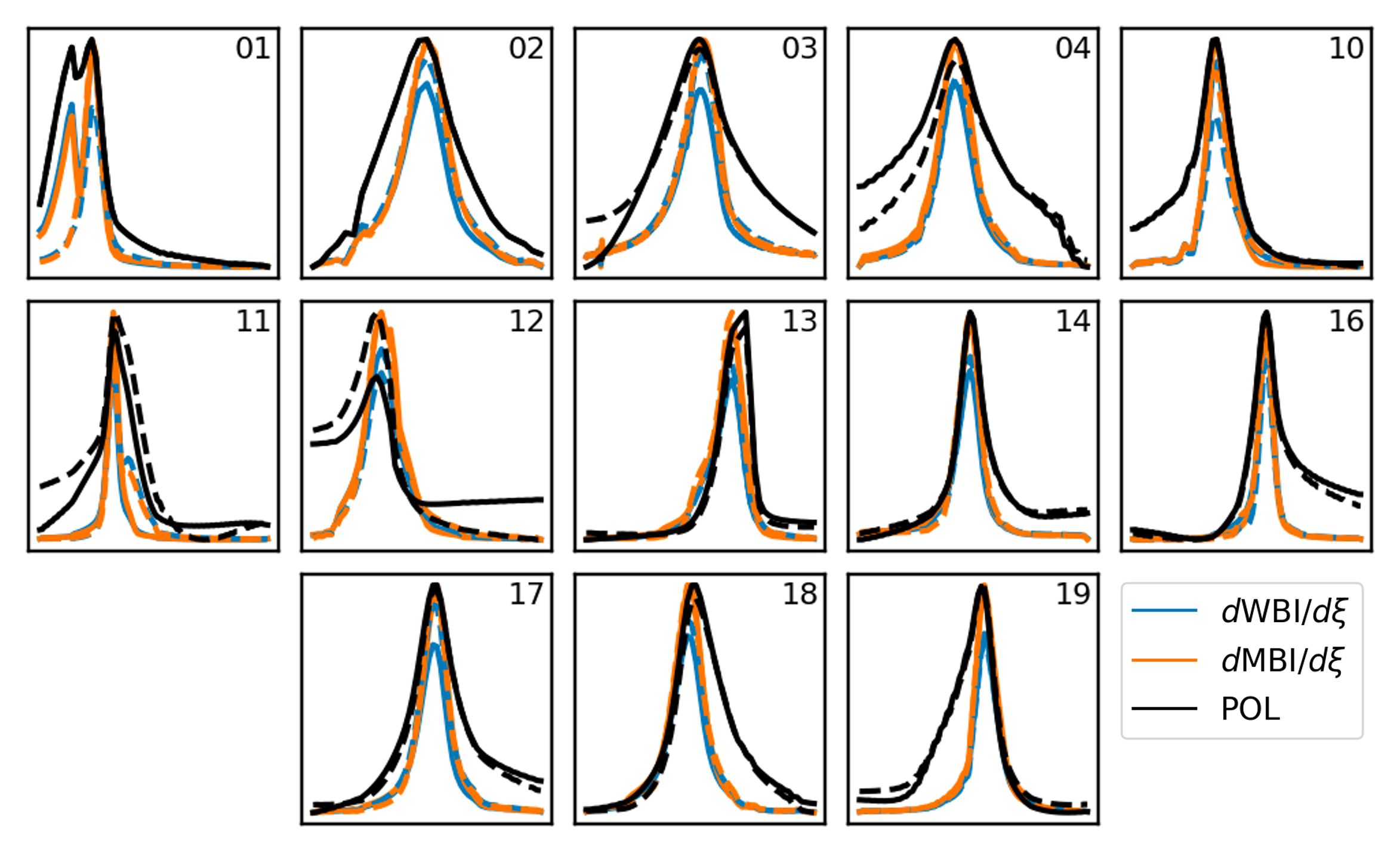}
    \caption{\textit{Bond order and bond order flux compared to parallel polarizability for 13 hydrogen combustion reactions}. The Wiberg (blue) and Mayer (orange) bond indices plotted alongside the bond projected polarizability (black). The plots are overlaid and are plotted on separate scales. Solid and broken lines indicate forming and breaking bonds, respectively.}
    \label{fig:bond-flux}
\end{figure}

\begin{figure}
    \centering
    \includegraphics[width=1\linewidth]{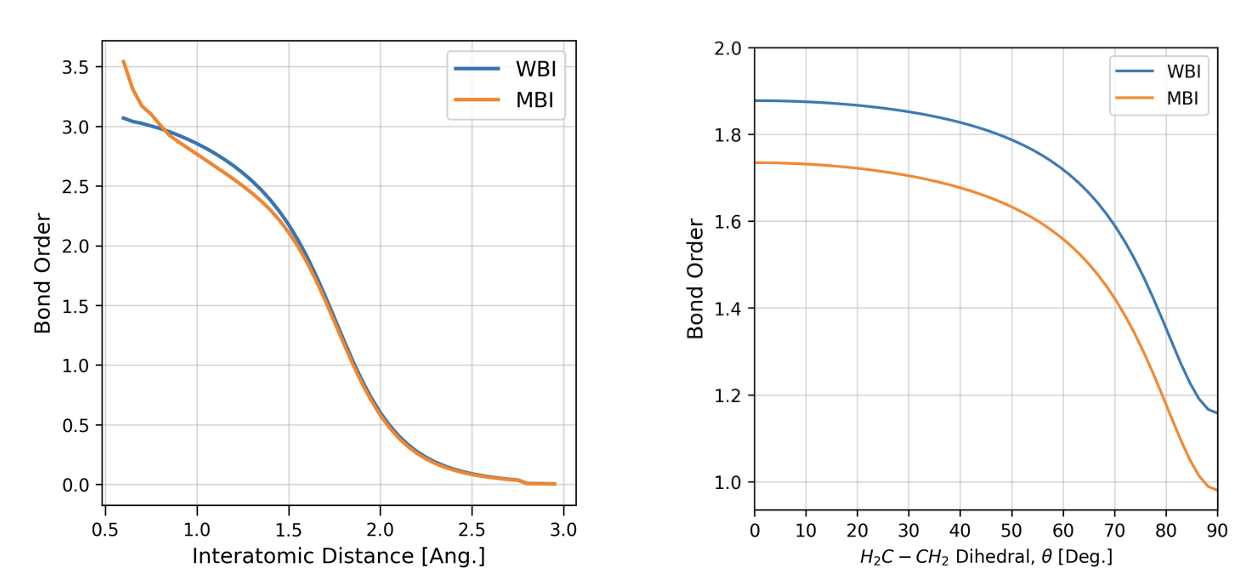}
    \caption{\textit{Nitrogen molecule dissociation and Ethene bond rotation} Bond order, bond order flux, and polarizability profiles computed with spin-unrestricted CASSCF(6,6)/cc-pVTZ and CASSCF(2,2)/cc-pVTZ for nitrogen and ethene, respectively.}
    \label{fig:nitrogen-dissociation}
\end{figure}